\documentclass[fleqn,usenatbib,letters]{mnras}
\usepackage{newtxtext,newtxmath}
\usepackage{graphicx}	
\usepackage{amsmath}	
\usepackage{amssymb}	

\usepackage{xspace} 

\newcommand{\XY}[2]{\left[\textrm{#1}/\textrm{#2}\right]}
\newcommand{\FeH}{\XY{Fe}{H}}
\newcommand{\XFe}[1]{\XY{#1}{Fe}}
\newcommand{\XH}[1]{\XY{#1}{H}}
\newcommand{\alphaFe}{\left[{\rm \alpha}/\textrm{Fe}\right]}

\newcommand{\kms}{km\,s$^{-1}$}
\newcommand{\Teff}{T_\textrm{eff}}
\newcommand{\logg}{\log g}

\newcommand{\vmic}{v_\textrm{mic}}
\newcommand{\Abund}[1]{A(\text{#1})}
\newcommand{\Msol}{\text M_\odot}

\newcommand{\thestar}{SMSS 1605$-$1443\@\xspace}

\newcommand*{\eg}{e.g.\@\xspace}
\newcommand*{\ie}{i.e.\@\xspace}

\usepackage{color}

\title[SMSS1605$-$1443]{The lowest detected stellar Fe abundance: The halo star SMSS J160540.18$-$144323.1}

\author[T. Nordlander et al.]{
T. Nordlander,$^{1,2}$\thanks{thomasn@mso.anu.edu.au}
M.S. Bessell,$^{1,2}$ 
G.S. Da Costa,$^1$ 
A.D. Mackey,$^{1,2}$  
M. Asplund,$^{1,2}$ \newauthor
A.R. Casey,$^{3,4}$ 
A. Chiti,$^5$ 
R. Ezzeddine,$^{5,6}$
A. Frebel,$^{5,6}$ 
K. Lind,$^{7,8}$ 
A.F. Marino,$^{1,9}$  \newauthor 
S.J. Murphy,$^{1,10}$
J.E. Norris,$^1$  
B.P. Schmidt,$^1$
and D. Yong,$^{1,2}$  
\\
$^1$Research School of Astronomy and Astrophysics, Australian National University, Canberra, ACT 2611, Australia\\
$^2$ARC Centre of Excellence for All Sky Astrophysics in 3 Dimensions (ASTRO 3D), Australia\\
$^3$School of Physics \& Astronomy, Monash University Clayton 3800, Victoria, Australia \\
$^4$Faculty of Information Technology, Monash University Clayton 3800, Victoria, Australia \\
$^5$Department of Physics \& Kavli Institute for Astrophysics and Space Research, Massachusetts Institute of Technology, Cambridge, MA 02139, USA \\
$^6$Joint Institute for Nuclear Astrophysics---Center for Evolution of the Elements, East Lansing, MI 48824, USA \\
$^7$Max-Planck-Institut f\"ur Astronomie, K\"onigstuhl 17, D-69117 Heidelberg, Germany \\
$^8$Observational Astrophysics, Department of Physics and Astronomy, Uppsala University, Box 516, 75120 Uppsala, Sweden \\
$^9$Dipartimento di Fisica e Astronomia Galileo Galilei, Univ. di Padova, Vicolo dellOsservatorio 3, Padova, IT-35122 \\
$^{10}$School of Science, University of New South Wales Canberra, ACT 2600, Australia
}

\date{Accepted XXX. Received YYY; in original form ZZZ}

\pubyear{2019}

\begin{document}
\label{firstpage}
\pagerange{\pageref{firstpage}--\pageref{lastpage}}
\maketitle

\begin{abstract}
We report the discovery of SMSS J160540.18$-$144323.1, a new ultra-metal poor halo star discovered with the SkyMapper telescope.
We measure $\FeH = -6.2 \pm 0.2$ (1D LTE), the lowest ever detected abundance of iron in a star.
The star is strongly carbon-enhanced, $\XFe{C} = 3.9 \pm 0.2$, while other abundances are compatible with an $\alpha$-enhanced solar-like pattern with $\XFe{Ca} = 0.4 \pm 0.2$, $\XFe{Mg} = 0.6 \pm 0.2$, $\XFe{Ti} = 0.8 \pm 0.2$, and no significant s- or r-process enrichment, $\XFe{Sr} < 0.2$ and $\XFe{Ba} < 1.0$ (3$\sigma$ limits). 
Population III stars exploding as fallback supernovae may explain both the strong carbon enhancement and the apparent lack of enhancement of odd-$Z$ and neutron-capture element abundances. 
Grids of supernova models computed for metal-free progenitor stars yield good matches for stars of about $10\,\Msol$ imparting a low kinetic energy on the supernova ejecta, while models for stars more massive than roughly $20\,\Msol$ are incompatible with the observed abundance pattern. 
\end{abstract}




\section{Introduction}

The early evolution of the Universe depends on the properties of the first generation of metal-free stars, the so-called Population III, and in particular on their mass as well as properties of their supernova explosions. 
High-mass Population III stars were short-lived, and can only be studied indirectly through their supernova ejecta that enriched the gas clouds from which the oldest metal-poor (but not metal-free) stars formed which are still observable today. 

Targeted efforts by several groups \citep[e.g.,][]{beers_search_1985,christlieb_finding_2003,keller_skymapper_2007,caffau_topos._2013,aguado_new_2017,starkenburg_pristine_2017} 
have led to the discovery of roughly 30 stars with $\FeH < -4$\footnote{Throughout this discussion we use the 1D LTE abundance values.} \citep{abohalima_jinabasedatabase_2018},
where the most iron-poor stars in fact only have upper limits. In particular, SMSS~0313$-$6708 at $\FeH < -7.3$ \citep[][]{keller_single_2014,nordlander_3d_2017} and J0023+0307 at $\FeH < -5.8$ \citep[][]{aguado_j0023+0307_2018,frebel_chemical_2019} both have abundance patterns that indicate true iron abundances (predicted from Population III star supernova models) significantly lower than their detection limits. The most iron-poor stars where iron has actually been detected are HE~1327$-$2326 at $\FeH = -5.7$ \citep[][]{frebel_nucleosynthetic_2005,aoki_he_2006}, HE~0107$-$5240 at $\FeH = -5.4$ \citep[][]{christlieb_stellar_2002,christlieb_he_2004}, and SD~1313$-$0019 at $\FeH = -5.0$ \citep[][]{allende_prieto_equatorial_2015,frebel_sd_2015}.
All five stars exhibit strong carbon enhancement and typically strong odd-even effects that are similar to predictions for Population III star supernovae with masses between 10 and 60\,$\Msol$, and explosion energies less than $10^{51}$\,erg assuming a mixing and fallback explosion mechanism \citep{heger_nucleosynthesis_2010,ishigaki_faint_2014}. 
In particular for the two stars that have only upper limits on their iron abundance, the comparison is not well constrained and matches instead for a wide range of progenitor mass and explosion energy \citep[][]{nordlander_3d_2017,frebel_chemical_2019}. 
This happens because the iron abundance is sensitive to processes that occur near the iron core of the progenitor star, \eg, the amount of mixing driven by Rayleigh-Taylor instabilities, the location where the explosion originates, and the explosion energy that determines whether ejecta subsequently fall back onto the newly formed black hole \citep[see discussion in][]{ishigaki_faint_2014}. 

We have recently discovered SMSS J160540.18$-$144323.1 (hereafter \thestar), a red giant branch star with the lowest ever detected abundance of iron, $\FeH = -6.2 \pm 0.2$. 
The fact that iron has been detected alongside carbon, magnesium, calcium and titanium, offers for the first time strong constraints on chemical enrichment at this metallicity.
We give here an assessment of its stellar parameters and chemical composition based on the spectra acquired during discovery and verification.

\section{Observations}
\thestar ($g=16.0$) was discovered as part of the SkyMapper search for extremely metal-poor stars \citep{keller_skymapper_2007,da_costa_skymapper_2019} using the metallicity-sensitive narrow-band $v$-filter in SkyMapper DR1.1 \citep{wolf_skymapper_2018}.
The star was confirmed to have $\FeH < -5$ from medium-resolution ($R=3000$ and $R=7000$) spectrophotometry acquired in March and August 2018 with the WIFES spectrograph \citep{dopita_wide_2010} on the ANU 2.3-metre telescope. The photometric selection and confirmation methodology is described further elsewhere \citep{jacobson_high-resolution_2015,marino_keck_2019,da_costa_skymapper_2019}. 
Follow-up high-resolution spectra were taken on the night of September 1 2018 in 1\,arcsec seeing with the MIKE spectrograph \citep{bernstein_MIKE_2003} at the 6.5m Magellan Clay telescope. We used a 1\,arcsec slit and 2x2 binning, producing a spectral resolving power $R = \lambda / \Delta \lambda = 28\,000$ on the blue detector and 22\,000 on the red detector. 
We reduced data using the CarPy pipeline \citep{kelson_optimal_2003}. Coadding the 4x1800\,s exposures resulted in a signal-to-noise per pixel, $S/N \approx 10$ at 3700\,\AA, 30 at 4000\,\AA, and 90 at 6700\,\AA.

\section{Methods}
We fit the observed high- and medium-resolution spectra using $\chi^2$ statistics. Upper limits to abundances were determined using a likelihood estimate assuming Gaussian errors, considering multiple lines simultaneously where applicable.
While fitting, the synthetic spectra are convolved with a Gaussian profile representing the instrumental profile.
We determine the continuum placement by taking the median ratio between the observed and synthetic spectrum in continuum windows that are predicted to be free of line absorption. 
In the spectrophotometric analysis, the slope of the continuum is matched by applying the $R_V = 3.1$ reddening law from \citet{fitzpatrick_correcting_1999} to the synthetic spectra.

For the spectrophotometric analysis, we compute a comprehensive grid of 1D LTE spectra using the Turbospectrum code \citep[v15.1;][]{alvarez_near-infrared_1998,plez_turbospectrum_2012} and MARCS model atmospheres \citep{gustafsson_grid_2008}.
We use $\vmic = 2$\,\kms\ and perform the radiative transport under spherical symmetry taking into account continuum scattering. The spectra are computed with a sampling step of 1\,\kms, corresponding to a resolving power $R \approx 300\,000$.
We adopt the solar chemical composition and isotopic ratios from \citet{asplund_chemical_2009}, but assume $\alphaFe = 0.4$ and compute spectra with varying carbon abundance. 
For our high-resolution spectroscopic abundance analyses, we compute additional grids where we vary the overall metallicity as well as the abundance of carbon and one additional element at a time.
We also use the 3D NLTE hydrogen Balmer line profiles from \citet{amarsi_effective_2018-1}. 

For all 1D LTE grids, we use a selection of atomic lines from VALD3 \citep{ryabchikova_major_2015} together with roughly 15 million molecular lines representing 18 different molecules, the most important of which for this work being those for CH \citep{masseron_ch_2014} and CN \citep{brooke_einstein_2014,sneden_line_2014}.

\section{Results}

\subsection{Stellar parameters}
\begin{figure*}
 \centerline{\includegraphics[width=\textwidth]{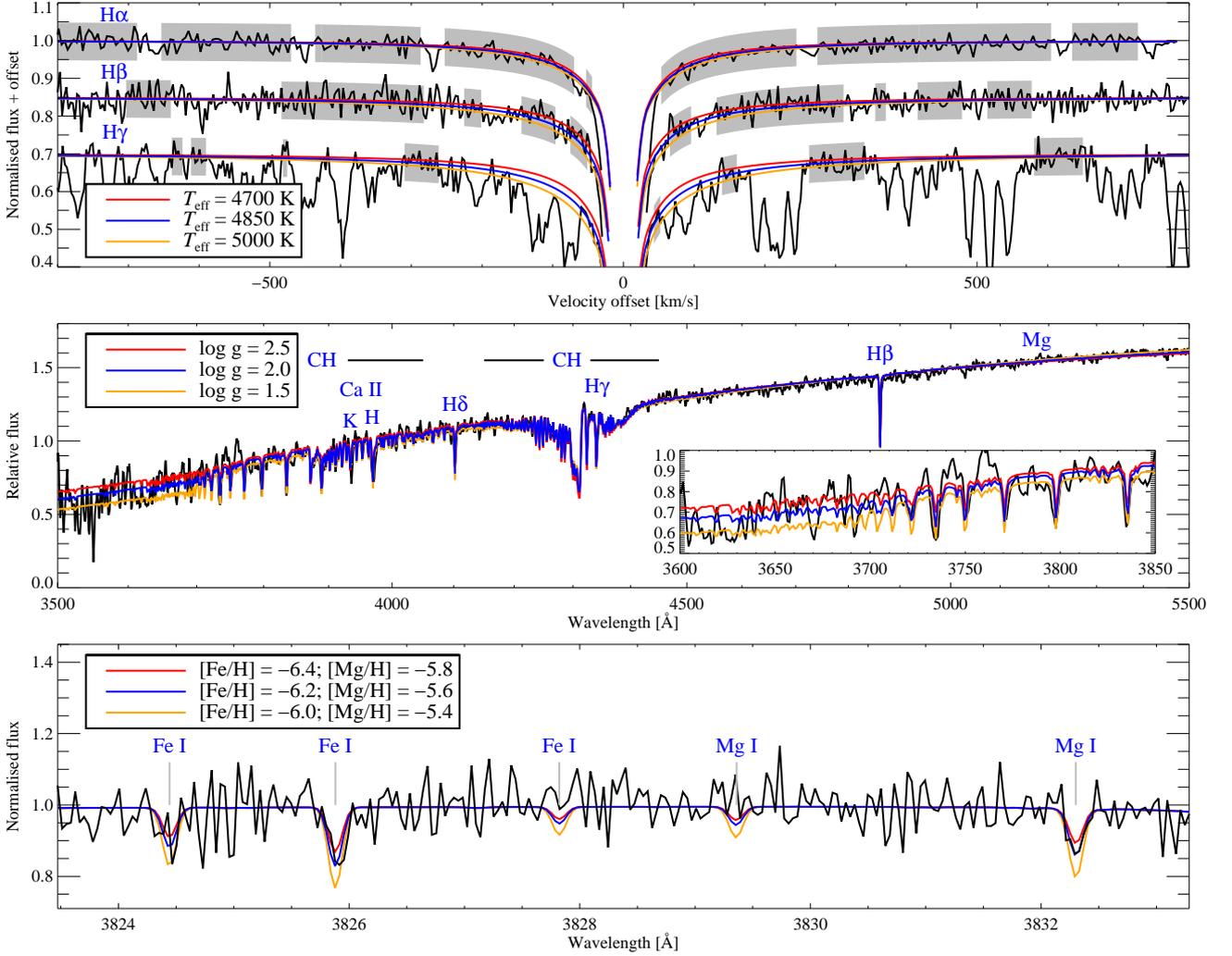}}
 \caption{
  \textit{Upper panel}: Fit of the effective temperature to the first three Balmer lines (labeled) in the MIKE high-resolution spectrum, compared to models at the preferred $\Teff = 4850$\,K. The lines are shown on a velocity scale centred on each line, and have been offset vertically. The grey shaded blocks represent the wavelength ranges used in the $\chi^2$ minimisation.
 \textit{Middle panel}: Fit of the surface gravity to the WiFeS medium-resolution spectrophotometry, with a zoomed inset showing the Balmer jump region, at the preferred $\logg = 2.0$. 
 \textit{Lower panel}: Example fits to lines of Fe and Mg in the MIKE high-resolution spectrum. 
 In all panels additional models illustrate the sensitivity, and the legend lists the models as shown from top to bottom. }
 \label{fig:Hlines}
\end{figure*}

We find consistent stellar parameters from medium-resolution spectrophotometry, optical and infrared photometry, high-resolution Balmer line analyses and stellar evolution constraints, and illustrate our synthetic spectrum fits in Fig.~\ref{fig:Hlines}.

Our spectrophotometric analysis of the initial medium-resolution spectrum indicates $\Teff = 4925$\,K, $\logg = 2.0$ and $\FeH < -4.75$ (see Da Costa et al. 2019). We assumed a reddening value $E(B-V) = 0.20$ based on the dust map from \citet[][rescaled according to \citealt{wolf_skymapper_2018}]{schlegel_maps_1998}. This is similar to the distance-dependent dust map of \citet{green_galactic_2018} that indicates $E(B-V) = 0.23 \pm 0.02$.
The interstellar lines of \ion{Na}ID 5890\,\AA\ and \ion{K}I 7699\,\AA\ show a complex structure of multiple components, indicating $E(B-V)$ between 0.12 and 0.21 using the calibrations of \citet{munari_equivalent_1997} and \citet{poznanski_empirical_2012}.
Adopting this range in reddening, we find good spectrophotometric fits for $\Teff = 4900 \pm 100$\,K, $\logg = 2.0 \pm 0.2$.
%
The infrared flux method calibrations on SkyMapper and 2MASS photometry from \citet{casagrande_skymapper_2019} indicate 
$\Teff = 4865 \pm 34 \pm 117$\,K from $g-K_s$ and $4784 \pm 59 \pm 83$\,K from $z-K_s$, where the error bars represent the uncertainties due to the measurement and reddening, respectively.

We fit 3D NLTE Balmer line profiles \citep{amarsi_effective_2018-1} to the high-resolution spectrum, taking care to avoid telluric lines for H$\alpha$ as well as lines of CH that contaminate H$\gamma$ and H$\beta$. We find good simultaneous fits for all three Balmer lines with $\Teff = 4850 \pm 100$\,K and $\logg = 2.0^{+0.5}_{-0.3}$.
These reddening-free estimates are in excellent agreement with the aforementioned spectrophotometric and photometric values, and we therefore adopt as our final parameters: $\Teff = 4850 \pm 100$\,K, $\logg = 2.0 \pm 0.2$\,dex. 
With these stellar parameters, the spectrophotometry indicates $E(B-V) = 0.12$, in agreement with the strengths of interstellar lines.

\citet{placco_carbon-enhanced_2014} present stellar evolution models that take into account varying enhancement of carbon and nitrogen. 
The fact that nitrogen is not detected in \thestar\ implies that the episode of extra mixing usually associated with thermohaline mixing \citep{eggleton_deep_2006,charbonnel_thermohaline_2007-1} has not yet occurred, and further that the surface abundance of carbon is not depleted ($<0.01$\,dex). 
This extra mixing episode is associated with significant theoretical uncertainty, both in the magnitude of effects and the evolutionary stage where they occur \citep{angelou_thermohaline_2011,henkel_phenomenological_2017,shetrone_constraining_2019-1}. 
Taking into account the systematic corrections discussed by \citet{placco_carbon-enhanced_2014}, our non-detection of nitrogen constrains $\logg > 1.9$, in agreement with our spectroscopic measurements.

The Gaia DR2 parallax measurement, $\pi = 0.0004 \pm 0.0544$\,mas \citep{brown_gaia_2018}, yields a lower limit to the distance to \thestar\ implying $\logg < 2.5$ ($3 \sigma$). 
Conversely, our spectroscopic estimate of $\logg = 2.0 \pm 0.2$ implies a predicted parallax of $\pi = 0.09 \pm 0.02$\,mas, \ie, a distance of $11 \pm 3$\,kpc, placing it on the other side of the Galaxy.
We note that its kinematics (with $v_\text{rad} = -224$\,\kms) indicate it being a normal inner halo star.

\subsection{Abundance analysis}
We report results of our abundance analysis in Table~\ref{table:abundances}, where statistical uncertainties on the absolute abundance are based on our $\chi^2$ analyses and upper limits are given at the $3 \sigma$ level. The systematic errors on the absolute abundances are estimated by changing the stellar parameters ($\Teff$, $\logg$, $\FeH$ and $\XH C$), one at a time according to their estimated uncertainty, and adding the effects in quadrature. 
We do not attempt to quantify the influence of hydrodynamic and non-LTE effects \citep[e.g.,][]{amarsi_non-lte_2016-1,nordlander_3d_2017}, but defer this to future work that incorporates a full 3D non-LTE analysis and higher-quality observations (Nordlander et al, \textit{in prep.}).

We estimate the iron abundance from a set of 16 lines of \ion{Fe}i, 10 detected and 6 upper limits, with lower excitation potential $E_\text{low}$ between 0 and 1.5\,eV. 
Using a maximum-likelihood estimate that also takes into account the 6 lines that have only upper limits, we find a mean abundance $\FeH = -6.21 \pm 0.17$, 
with a flat trend $-0.01 \pm 0.14\,\text{dex}\,\text{eV}^{-1}$ as a function of $E_\text{low}$. 
\ion{Fe}{ii} cannot be detected using the current spectrum. The three strongest lines yield an upper limit $\FeH < -4.7$ ($3\sigma$).

We estimate a carbon abundance $\XH{C} = -2.32 \pm 0.05$ using CH lines from the $A^2\Delta$--$X^2\Pi$ system at 4100--4400\,\AA\ and the $B^2\Sigma^-$--$X^2\Pi$ system at 3900\,\AA. 
We do not detect absorption due to $^{13}$CH, and refrain from placing a limit on the isotopic ratio.
For magnesium, we measure $\XH{Mg} = -5.65 \pm 0.13$ from the UV triplet at 3829--3838\,\AA. We find an equivalent width of just 17\,m\AA\ for the only detectable \ion{Mg}ib line at 5185\,\AA.
For calcium, the \ion{Ca}{ii} H and K lines indicate $\XH{Ca} = -5.07 \pm 0.05$. We also measure $\XH{Ca} = -5.85 \pm 0.11$ from \ion{Ca}{i} 4226\,\AA, resulting in a very large 0.8\,dex abundance difference between the two ionisation stages. This is likely mainly due to the non-LTE overionisation of \ion{Ca}i as well as a smaller non-LTE effect of opposite sign acting on \ion{Ca}{ii} \citep[see \eg,][]{sitnova_ultra_2019}. 
Comparing the measured abundances of \ion{Ca}i and \ion{Fe}i, this implies a normal level of $\alpha$-enhancement as seen in most halo stars, $\XY{Ca}{Fe} = 0.37 \pm 0.20$.
For titanium we detect the two lines of \ion{Ti}{ii} at 3759--3761\,\AA\ and obtain $\XH{Ti} = -5.40 \pm 0.10$.

We determine upper limits for additional elements using a likelihood estimate that assumes Gaussian errors. We use synthetic spectra for these estimates, and consider multiple lines simultaneously when applicable.

\begin{table}
 \caption{High-resolution spectroscopic 1D LTE abundance analysis. Upper limits are given at the $3 \sigma$ level. Error estimates on the absolute abundances are reported for both the statistical measurement uncertainty ($\sigma_\text{stat}$) and the systematic uncertainty due to uncertainties in stellar parameters ($\sigma_\text{sys}$). The last column gives the reference solar chemical composition.}
 \label{table:abundances}
 \begin{tabular}{l rrr cc c}
 \hline
          Species      &    $\Abund{X}$ &   $\XH{X}$ & $\XFe{X}$ & $\sigma_\text{stat}$ & $\sigma_\text{sys}$ &    $\Abund{X}_{\odot}$ \\ 
 \hline
          \ion{Li}i    & $<  0.48$ & $< -0.57$ & $<  5.64$  &   0.18 &   0.09 &   1.05 \\
          C (CH)       & $   6.07$ & $  -2.32$ & $   3.89$  &   0.05 &   0.27 &   8.39 \\
          N (CN)       & $<  4.80$ & $< -2.98$ & $<  3.23$  &   0.19 &   0.18 &   7.78 \\
          \ion{O}i     & $<  7.21$ & $< -1.48$ & $<  4.73$  &   0.19 &   0.15 &   8.69 \\
          \ion{Na}i    & $<  0.90$ & $< -5.27$ & $<  0.94$  &   0.18 &   0.10 &   6.17 \\
          \ion{Mg}i    & $   1.88$ & $  -5.65$ & $   0.57$  &   0.13 &   0.09 &   7.53 \\
          \ion{Al}i    & $<  0.67$ & $< -5.76$ & $<  0.45$  &   0.19 &   0.11 &   6.43 \\
          \ion{Si}i    & $<  2.09$ & $< -5.42$ & $<  0.80$  &   0.20 &   0.11 &   7.51 \\
          \ion{K}i     & $<  1.98$ & $< -3.10$ & $<  3.11$  &   0.19 &   0.09 &   5.08 \\
          \ion{Ca}i    & $   0.46$ & $  -5.85$ & $   0.37$  &   0.11 &   0.13 &   6.31 \\
          \ion{Ca}{ii} & $   1.24$ & $  -5.07$ & $   1.15$  &   0.05 &   0.15 &   6.31 \\
          \ion{Sc}{ii} & $< -1.76$ & $< -4.93$ & $<  1.29$  &   0.12 &   0.10 &   3.17 \\
          \ion{Ti}{ii} & $  -0.50$ & $  -5.40$ & $   0.82$  &   0.10 &   0.10 &   4.90 \\
          \ion{V}{ii}  & $<  0.69$ & $< -3.31$ & $<  2.90$  &   0.23 &   0.09 &   4.00 \\
          \ion{Cr}i    & $<  0.22$ & $< -5.42$ & $<  0.79$  &   0.20 &   0.13 &   5.64 \\
          \ion{Mn}i    & $<  0.03$ & $< -5.36$ & $<  0.85$  &   0.19 &   0.15 &   5.39 \\
          \ion{Fe}i    & $   1.24$ & $  -6.21$ &     \dots  &   0.17 &   0.14 &   7.45 \\
          \ion{Fe}{ii} & $<  2.72$ & $< -4.73$ &     \dots  &   0.18 &   0.06 &   7.45 \\
          \ion{Co}i    & $<  0.56$ & $< -4.36$ & $<  1.85$  &   0.19 &   0.14 &   4.92 \\
          \ion{Ni}i    & $<  0.87$ & $< -5.36$ & $<  0.85$  &   0.25 &   0.14 &   6.23 \\
          \ion{Cu}i    & $<  1.51$ & $< -2.70$ & $<  3.51$  &   0.19 &   0.12 &   4.21 \\
          \ion{Zn}i    & $<  1.55$ & $< -3.05$ & $<  3.16$  &   0.19 &   0.06 &   4.60 \\
          \ion{Sr}{ii} & $< -3.12$ & $< -6.04$ & $<  0.17$  &   0.19 &   0.10 &   2.92 \\
          \ion{Ba}{ii} & $< -3.07$ & $< -5.24$ & $<  0.97$  &   0.19 &   0.11 &   2.17 \\
          \ion{Eu}{ii} & $< -2.41$ & $< -2.93$ & $<  3.28$  &   0.19 &   0.11 &   0.52 \\        
 \hline
 \end{tabular}
\end{table}

\section{Discussion}
\begin{figure}
 \centerline{\includegraphics[]{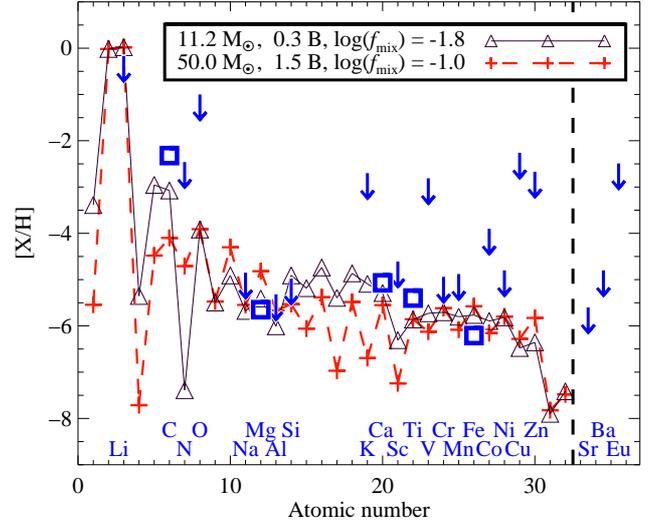}}
 \caption{Comparison of the measured abundances (blue squares) and $3 \sigma$ upper limits (blue arrows) to yields from Pop III supernova models with varying progenitor star mass, explosion energy (in units of $10^{51}\,\text{erg} = 1\,\text{B}$), and mixing parameter. No satisfactory fit to the observed abundance pattern exists for models more massive than about $20\,\Msol$.
\label{fig:sne}
 }
\end{figure}

Our analysis of \thestar reveals remarkably low abundances of heavier elements, including the lowest ever measured abundance of iron at $\FeH = -6.2 \pm 0.2$. 
While the abundance pattern from Na to Zn is broadly compatible with a standard $\alpha$-enhanced chemical composition typical of halo stars, the large carbon enhancement is a strong indicator for enrichment from a Population III mixing-and-fallback supernova \citep[see \eg][]{umeda_nucleosynthesis_2002,nomoto_nucleosynthesis_2013}. 
Using the predicted supernova yields computed for metal-free Population III stars by \citet{heger_nucleosynthesis_2010}, we find a reasonable match only for 
low-mass progenitors ($M \approx 10\,\Msol$) with low explosion energy ($<10^{51}$\,erg), 
as shown in Fig.~\ref{fig:sne}. Models more massive than about $20\,\Msol$ cannot simultaneously reproduce the strong carbon enhancement and the otherwise flat abundance trend.

Alternative explanations are unsatisfactory. 
The elevated abundance of carbon could be due to pollution from an intermediate-mass companion star, but models predict that this also leads to similar enhancement of nitrogen and neutron-capture elements \citep{campbell_evolution_2008,campbell_evolution_2010,cruz_s-process_2013}. 
An initially metal-free, or perhaps metal-poor but carbon-normal, star could also be polluted by accretion from the ISM. Again, models of this process predict significant enhancement of nitrogen alongside carbon relative to the depletion of refractory iron-peak elements \citep{johnson_chemical_2015}, and can likewise be ruled out.

It has been shown in previous work \citep{collet_chemical_2006,frebel_he_2008,caffau_primordial_2012,bessell_nucleosynthesis_2015,nordlander_3d_2017} that significant systematic uncertainties are associated with the chemical abundance analyses of the most iron-poor stars. We note that these corrections depend sensitively on not only the effective temperature and surface gravity of the star, but also the abundance of the element under study, and we caution against blindly applying representative corrections. 
Although these effects may be as large as 1\,dex, they are unlikely to significantly alter the main conclusions of this work: It is clear that \thestar is the most iron-deficient star for which iron has been detected, that it is strongly carbon enhanced, and that it does not exhibit strong enhancement nor a strong abundance trend among elements heavier than carbon. 
A higher-quality spectrum would enable more stringent limits and likely detections of additional elements, which together with advanced spectrum synthesis techniques will allow us to better understand the properties of the Pop III progenitor star.

\section*{Acknowledgements}
We thank Richard Stancliffe for providing data on the carbon-enhanced stellar evolution models used in this work.

Parts of this research were conducted by the Australian Research Council Centre of Excellence for All Sky Astrophysics in 3 Dimensions (ASTRO 3D), through project number CE170100013.
Research on extremely metal-poor stars has been supported in part through Australian Research Council Discovery Grant Program DP150103294 (G.S.D.C., M.S.B. and B.P.S.).
A.R.C. is supported in part by Australian Research Council Discovery Project DP160100637.
A.D.M. is supported by an Australian Research Council Future Fellowship (FT160100206).

The national facility capability for SkyMapper has been funded through ARC LIEF grant LE130100104 from the Australian Research Council, awarded to the University of Sydney, the Australian National University, Swinburne University of Technology, the University of Queensland, the University of Western Australia, the University of Melbourne, Curtin University of Technology, Monash University and the Australian Astronomical Observatory. SkyMapper is owned and operated by The Australian National University's Research School of Astronomy and Astrophysics. The survey data were processed and provided by the SkyMapper Team at ANU. The SkyMapper node of the All-Sky Virtual Observatory (ASVO) is hosted at the National Computational Infrastructure (NCI). Development and support the SkyMapper node of the ASVO has been funded in part by Astronomy Australia Limited (AAL) and the Australian Government through the Commonwealth's Education Investment Fund (EIF) and National Collaborative Research Infrastructure Strategy (NCRIS), particularly the National eResearch Collaboration Tools and Resources (NeCTAR) and the Australian National Data Service Projects (ANDS).

We also acknowledge the traditional owners of the land on which the SkyMapper telescope stands, the Gamilaraay people, and pay our respects to elders past, present and emerging.



\bibliographystyle{mnras}
\bibliography{references}





\bsp	
\label{lastpage}
\end{document}